\documentclass{PoS}

\title{The strange and charm quark contributions to the anomalous magnetic moment (g -2) of the muon from current-current correlators}

\ShortTitle{The strange and charm quark contributions to the muon anomaly (g-2)}

\author{\speaker{Bipasha Chakraborty}\\SUPA, School of Physics and Astronomy, University of Glasgow, Glasgow G12 8QQ, UK\\
        E-mail: \email{b.chakraborty.1@research.gla.ac.uk}}
\author{Christine Davies\\SUPA, School of Physics and Astronomy, University of Glasgow, Glasgow G12 8QQ, UK}
\author{Gordon Donald\\Institut f\"{u}r Theoretische Physik, Universit\"{a}t Regensburg, 93040 Regensburg, Germany}
\author{Rachel Dowdall\\DAMTP, University of Cambridge, Wilberforce Road, Cambridge CB3 0WA, UK}
\author{Pedro Gon\c{c}alves de Oliveira\\SUPA, School of Physics and Astronomy, University of Glasgow, Glasgow G12 8QQ, UK}
\author{Jonna Koponen\\SUPA, School of Physics and Astronomy, University of Glasgow, Glasgow G12 8QQ, UK}
\author{G. Peter Lepage\\Laboratory of Elementary Particle Physics, Cornell University, Ithaca, NY 14853, USA}
\author{T. Teubner\\Department of Mathematical Sciences, University of Liverpool, Liverpool, L69 3BX, UK}
                              
\abstract{We describe a new technique (presented in arXiv:1403.1778) to determine the contribution to the anomalous magnetic moment (g-2) of the muon coming from the hadronic vacuum polarization using lattice QCD. Our method uses Pad\'{e} approximants to reconstruct the Adler function from its derivatives at $q^2=0$. These are obtained simply and accurately from time-moments of the vector current-current correlator at zero spatial momentum. We test the method using strange quark correlators calculated on MILC Collaboration's $n_f$ = 2+1+1 HISQ ensembles at multiple values of the lattice spacing, multiple volumes and multiple light sea quark masses (including physical pion mass configurations).}

\FullConference{The 32nd International Symposium on Lattice Field Theory\\
                 23-28 June, 2014\\
                 Columbia University New York, NY}

\begin{document}

\section{Motivation} 
\label{sec:intro2}

\begin{figure}
\centering
\includegraphics[width=0.28\textwidth]{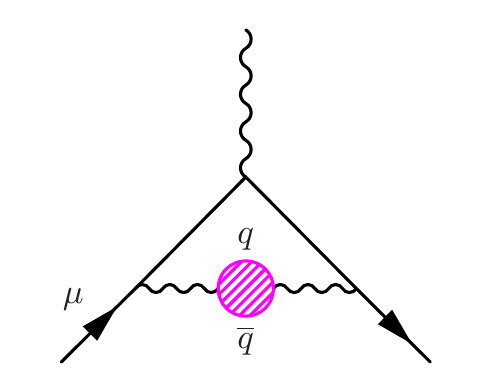}
\caption{The hadronic vacuum polarization contribution to the muon anomalous magnetic moment is represented as a shaded blob inserted into the photon propagator (represented by a wavy line) that corrects the point-like photon-muon coupling at the top of the diagram.}
\label{fig:hvp}
\end{figure}
The anomalous magnetic moment of the muon or the muon anomaly ($a_\mu$) has been measured with very impressive accuracy of 0.54 ppm~\cite{Bennett:2006fi} in the experiment (BNL E821), therefore providing one of the most stringent tests of the Standard Model. The anomaly, defined as the fractional difference of its gyromagnetic ratio from the naive value of 2, ($a_{\mu}=(g-2)/2$), arises from the muon interacting with a cloud of virtual particles and gets contributions from QED, electroweak (EW) and QCD (hadronic loops) diagrams. Intriguingly, the experimentally measured anomaly disagrees by around 3$\sigma$ with the calculated value from the Standard Model : $a_\mu^{exp}-a_\mu^{SM}=25(9) \times 10^{-10}$ ~\cite{Aoyama:2012wk, Hagiwara:2011af,rafael1972, Davier}. The current theoretical uncertainty is dominated by that from the theoretical calculation of the lowest order ``hadronic vacuum polarisation (HVP)'', $a_{\mu}^{HVP, LO}$. This contribution is currently determined most accurately using the dispersion relation and the experimental results on $e^+e^- \rightarrow$ hadrons or from $\tau$ decay to be of size $700 \times 10^{-10}$ with $\sim$1$\%$ error~\cite{Hagiwara:2011af, Davier}. With a further goal to improve the experimental uncertainty by a factor of 4 in the upcoming experiment at Fermilab (E989), improvement in the theoretical determination, mainly by achieving an uncertainty of less than 1$\%$ in $a_{\mu}^{HVP,LO}$ using a first principle lattice QCD calculation is therefore essential in trying to ascertain the possibility of new physics (if the discrepancy still remains). We have developed a simple lattice QCD method~\cite{PhysRevD.89.114501} for calculating $a_{\mu}^{HVP,LO}$ which improves significantly on previous calculations.

\section{Our Method }

\begin{figure}
\centering
   \includegraphics[width=4.0cm,angle=0]{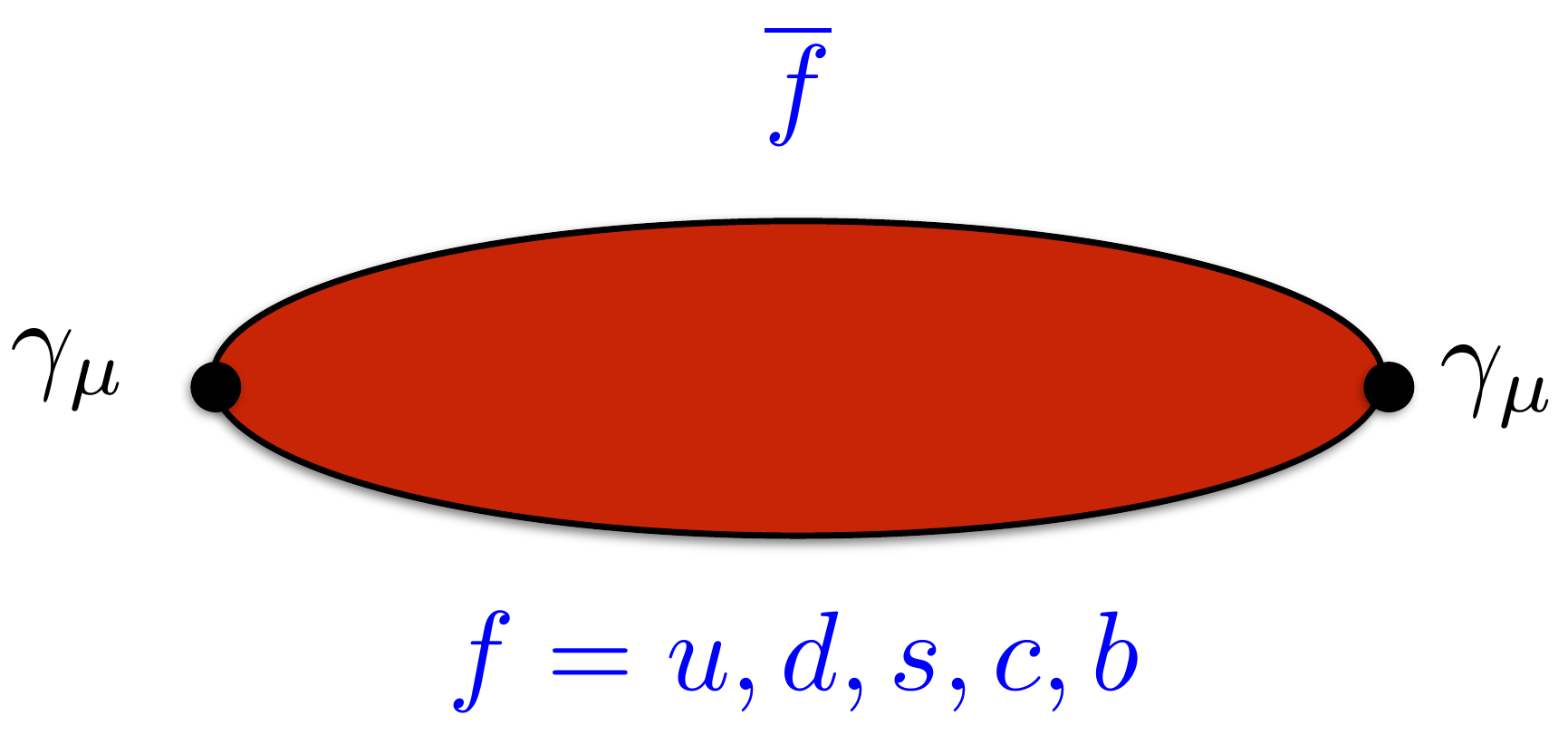}
\caption{The quark vacuum polarisation for each of the quark flavor f (represented by the shaded blob in figure 1).}
\label{fig:blob}
\end{figure} 
%\begin{equation}
%a_{\mu, \mathrm{HVP}}^{(\mathrm{f})} = \frac{\alpha}{\pi} \int_0^{\infty} dq^2 f(q^2) (4\pi\alpha Q_{\mathrm{f}}^2) \hat{\Pi}_{\mathrm{f}}(q^2)  
%\label{eq:amu}
%\end{equation}

The contribution to the muon anomalous magnetic moment from the HVP associated with a given quark flavour, $\mathrm{f}$, is obtained by inserting the quark vacuum polarisation (figure~\ref{fig:blob}) into the photon propagator~\cite{rafael1994, Blum:2002ii} : $a_{\mu, \mathrm{HVP}}^{(\mathrm{f})} = \frac{\alpha}{\pi} \int_0^{\infty} dq^2 f(q^2) (4\pi\alpha Q_{\mathrm{f}}^2) \hat{\Pi}_{\mathrm{f}}(q^2)$. 

Here, $\alpha \equiv \alpha_{\mathrm{QED}}$ and $Q_{\mathrm{f}}$ is the electric charge of quark $\mathrm{f}$ in units of $e$, $f(q^2)$ is a known function of the four momentum squared strongly peaked at very small $q^2$ values, $q^2 \sim m_{\mu}^2/4 \sim$ 0.003GeV$^2$, meaning that we need $\hat{\Pi}(q^2)$ for small $q^2$ values. 
%o as the integrand in  $a_{\mu}$. 
%Calculating directly at small $q^2$ using twisted boundary condition produces noisy results, whereas 
%Extrapolation from higher values of $q^2$, however, leads to large model uncertainties. 
Our method reconstructs $\hat{\Pi}$ from its derivatives at $q^2=$0, which can be simply and accurately calculated from time moments of local-local vector current correlators at zero spatial momentum. In the present calculation we have only considered the quark-line-connected contributions to the HVP since the disconnected pieces are suppressed by quark mass factors. 

The quark polarisation tensor is the Fourier transform of the vector current-current correlator. For spatial currents at zero spatial momentum 
\begin{equation}
\Pi^{ii}(q^2) =
q^2{\Pi}(q^2) = a^4 \sum_t e^{iqt} \sum_{\vec{x}}\langle j^{i}(\vec{x},t)j^{i}(0) \rangle. 
\label{eq:pi}
\end{equation}
We need the renormalised vacuum polarisation function, $\hat\Pi(q^2)\equiv\Pi(q^2)-\Pi(0)$. 

Time-moments of the correlator give the derivatives at $q^2=0$ of $\hat{\Pi}$ (see, for example, \cite{Allison:2008xk,McNeile:2010ji}): 
\vspace{-1em}\begin{eqnarray}
G_{2n} &\equiv& a^4 
\sum_t \sum_{\vec{x}} t^{2n} Z_V^2 \langle j^{i}(\vec{x},t)j^{i}(0) \rangle  
= (-1)^n \left. \frac{\partial^{2n}}{\partial q^{2n}} q^2\hat{\Pi}(q^2) \right|_{q^2=0} .
\label{eq:G}
\end{eqnarray}
Here we have allowed for a renormalisation factor $Z_V$ for the lattice vector current. We have evaluated $G_{2n}$ from the correlators calculated in lattice QCD for n$=$4,6,8,10, remembering that time runs from 0 at the origin in both positive and negative directions to a value of $\pm T/2$ in the centre of the lattice. 

We define a Taylor series expansion as : $\hat{\Pi}(q^2) = \sum_{j=1}^{\infty} q^{2j} \Pi_j$, then $\Pi_j = (-1)^{j+1} \frac{G_{2j+2}}{(2j+2)!}$.
%\begin{equation}
%\hat{\Pi}(q^2) = \sum_{j=1}^{\infty} q^{2j} \Pi_j  
%\label{eq:pihat}
%\end{equation}

%and the Taylor coefficients as : $\Pi_j = (-1)^{j+1} \frac{G_{2j+2}}{(2j+2)!}$.
%\begin{equation}
%\Pi_j = (-1)^{j+1} \frac{G_{2j+2}}{(2j+2)!} \, .
%\label{eq:derivs}
%\end{equation}

To evaluate the contribution to $a_{\mu}$ we replace $\hat{\Pi}(q^2)$ with its $[2,2]$ Pad\'{e} approximants derived from the~$\Pi_j$~\cite{Pade}. Using Pad\'{e} approximants allows us to obtain a sensible behaviour at high $q^2$ values. The [N, N] and [N, N+1] approximants bracket the exact result and converge to it as N increases~\cite{PhysRevD.89.114501}. The precision of different Pad\'{e} approximants was tested by comparing with the exact one-loop perturbative results for $a_{\mu}$~\cite{PhysRevD.89.114501}. We perform the $q^2$ integral numerically. 

\section{Simulation Details}
We calculate the strange ($s$) quark propagators using the Highly Improved Staggered Quark (HISQ)~\cite{HISQ_PRD} discretisation on HISQ gauge configurations generated by the MILC collaboration~\cite{Bazavov:2010ru, Bazavov:2012uw} with light (up/down), strange and charm quarks in the sea. We have used three lattice ensembles with lattice spacings a $\approx$ 0.15\,fm (verycoarse), 0.12\,fm (coarse) and 0.09\,fm (fine), determined~\cite{fkpi} using the Wilson flow parameter $w_0$~\cite{Borsanyi:2012zs}. At each lattice spacing we have two values of the average $u/d$ quark mass: one fifth the $s$ quark mass and the physical value ($m_s/27.5$). We tune the valence $s$ quark mass accurately~\cite{Chakraborty:2014zma} using the mass of the $\eta_s$ meson (688.5(2.2)\,MeV)~\cite{fkpi}. At a third value of the $u/d$ quark mass, one tenth of the $s$ quark  mass, at one lattice spacing $\sim$ 0.12\,fm we have three different volumes to test for finite volume effects. These sets correspond to a lattice length in units of the $\pi$ meson mass of $M_{\pi}L=$ 3.2, 4.3 and 5.4. In addition we de-tuned the valence $s$ quark mass there by 5$\%$ (set 6) to test for tuning effects.

The $s$ quark propagators are combined into a correlator with a local vector current at either end to form the vector meson $\phi$. The end point is summed over spatial sites on a timeslice to set the spatial momentum to zero. We use the random colour wall source created from a set of U(1) random numbers over a timeslice for improved statistics. The local current is not the conserved vector current for the HISQ quark action and must be renormalised. We have found the local vector current renormalisation constant ($Z_{V,\overline{s}s}$) completely non-perturbatively with 0.1$\%$ uncertainty on the finest $m_l=m_s/5$ lattices\cite{Chakraborty:2014zma}.

\section{Our results}

\subsection{Properties of $\phi$ meson}

\begin{figure}
   \centering                                                                                                                      
   \includegraphics[width=7.5cm,angle=0]{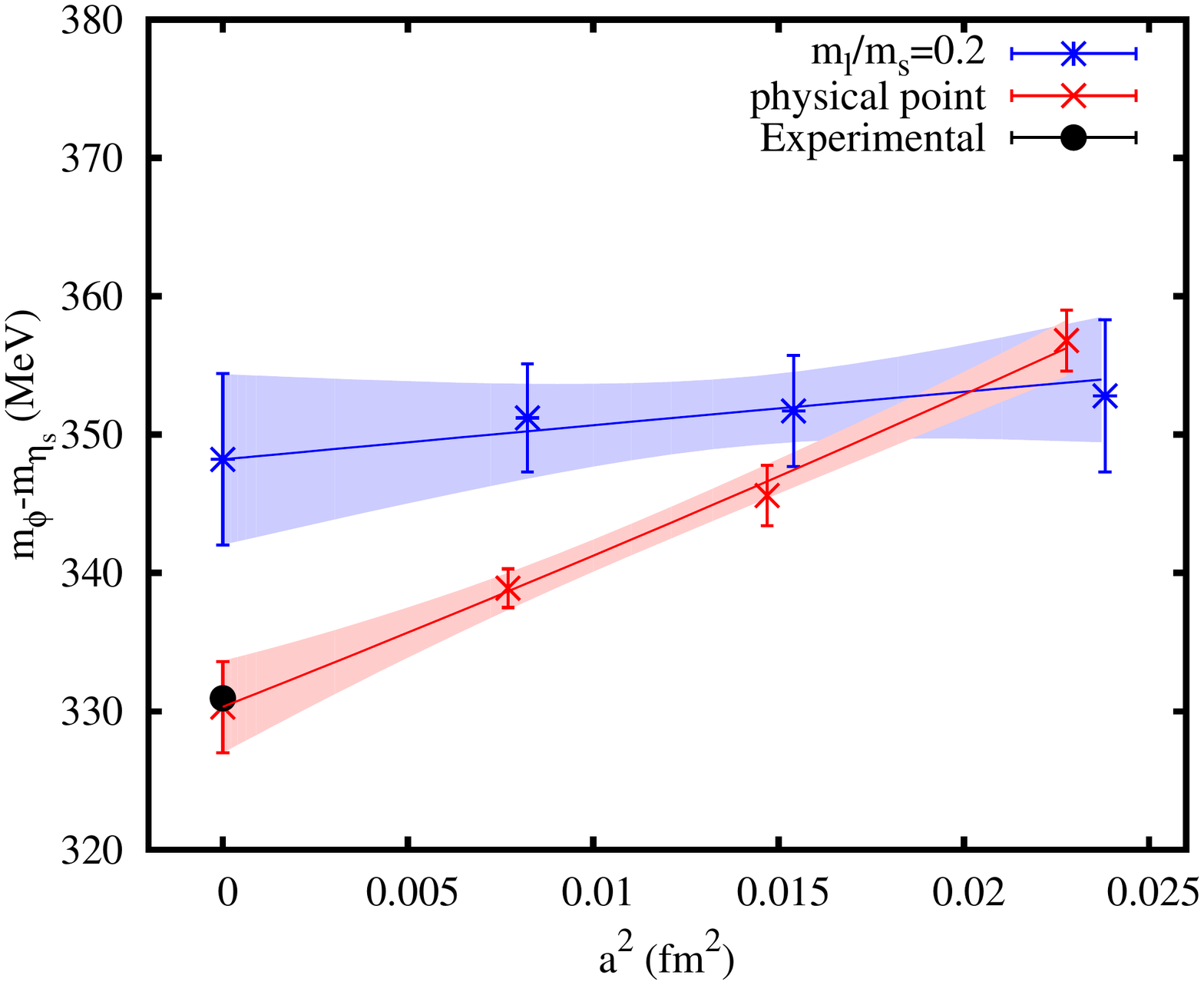}
   \hspace{-2em}
   \includegraphics[width=7.5cm,angle=0]{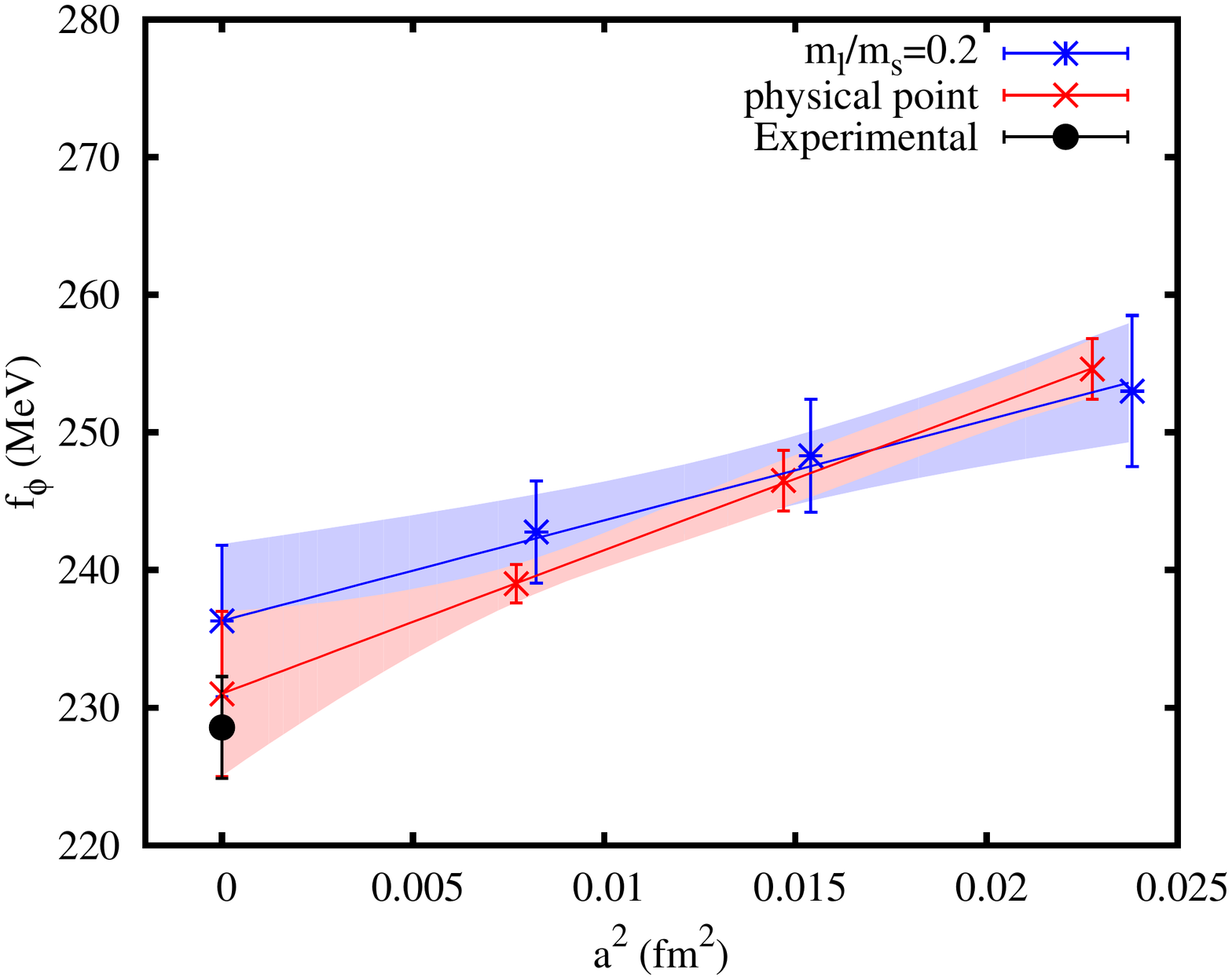}
   \caption{The plot on the left represents the results for $m_{\phi}-m_{{\eta}_s}$ calculated using the HISQ formalism on $m_l=m_s/5$ and physical point ensembles with varied lattice spacings and extrapolated to $a=0$. The continuum results are compared to the experiment. The plot on the right shows the similar results for $f_{\phi}$, compared to the experimental result derived from $\Gamma(\phi \rightarrow e^+e^-)$.}
\label{fig:phi}                                                                    
\end{figure}

We are concerned with the properties of the correlation function at the shorter times that feed into the theoretical determination of~$a_{{\mu},\mathrm{HVP}}$. But at large time separations between source and sink the correlators give the mass ($m_{\phi}$) and decay constant ($f_{\phi}$) of the $\phi$ meson~\cite{Chakraborty:2014zma}. The plots in figure~\ref{fig:phi} show how precisely we can extract those properties of the $\phi$ meson, and therefore, how accurate our correlators are. Our results for $m_\phi-m_{\eta_s}$ and $f_\phi$ in the continuum limit on the physical point lattices agree with experimental results. $f_{\phi}$ is related to $\Gamma(\phi \rightarrow e^+e^-)$.
%Finite volume effects seemed to be negligibly small. But, the valence HISQ strange quark mass tuning effect was significant. Disconnected diagrams are not included in the calculation, but we expect these to have only a very small effect.  

\subsection{Connected contributions to $a_\mu^s$ from full LQCD}

\begin{figure}
   \centering
    \includegraphics[width=0.5\textwidth]{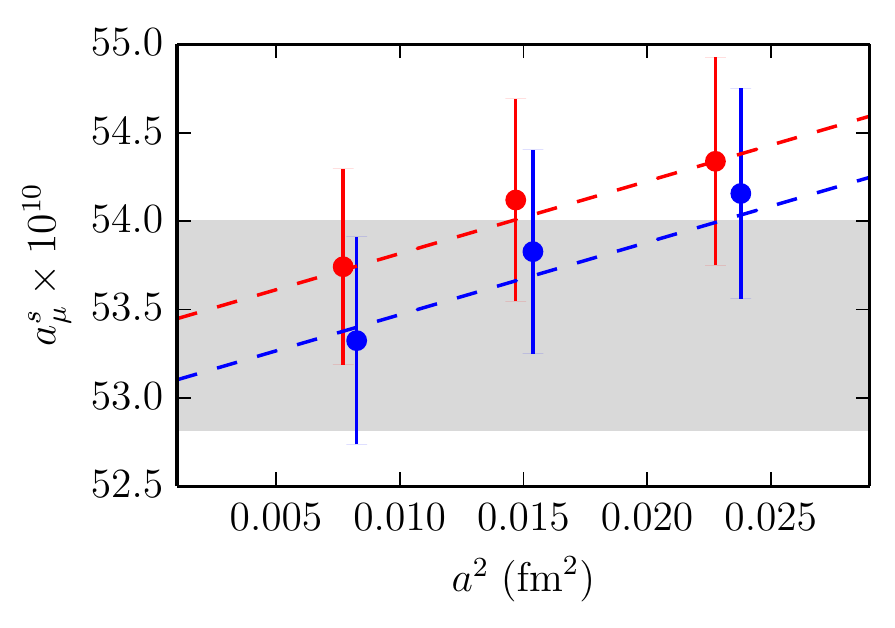}
    \caption{Lattice QCD results for the connected contribution to 
    the muon anomaly $a_\mu$ from vacuum polarisation of $s$~quarks. 
    Results are for three lattice
    spacings, and two light-quark masses:
    $m_\ell^\mathrm{lat} = m_s/5$ (lower, blue points), 
    and $m_\ell^\mathrm{lat} = m_\ell^\mathrm{phys}$ (upper, red
    points). The dashed lines are the corresponding values
    from the fit function,using the best-fit parameters. 
    The gray band shows our final result, $53.41(59)\times10^{-10}$, 
    with $m_\ell^\mathrm{lat}=m_\ell^\mathrm{phys}$,
    after extrapolation to $a=0$.
    }
    \label{fig:s-fit}
\end{figure}
We fit the results of $a_\mu^s$ using $[2,2]$ Pad\'{e} approximants from each configuration set to a function of the form
\begin{eqnarray}
      a^s_{\mu,\mathrm{lat}} &=& a^s_{\mu} \times \left( 1 +
     %c_{a^2}\left(\frac{a\Lambda_\mathrm{QCD}}{\pi} \right)^2 
       c_{a^2} (a\Lambda_\mathrm{QCD}/\pi)^2
      + c_\mathrm{sea} \delta x_\mathrm{sea} + c_\mathrm{val}\delta x_\mathrm{val}
      \right),\nonumber
  \end{eqnarray}
where $\Lambda_\mathrm{QCD}=0.5$\,GeV and \hspace{1em} $\delta x_\mathrm{sea} \equiv    \sum_{q=u,d,s}    \frac{m_{q}^\mathrm{sea} - m_{q}^\mathrm{phys}}{m_{s}^\mathrm{phys}}$,\hspace{2em} $\delta x_s \equiv\frac{m_{s}^\mathrm{val} - m_{s}^\mathrm{phys}}{m_{s}^\mathrm{phys}}$. 
%\begin{eqnarray}
%    \delta x_\mathrm{sea} &\equiv&
%    \sum_{q=u,d,s}
 %   \frac{m_{q}^\mathrm{sea} - m_{q}^\mathrm{phys}}{m_{s}^\mathrm{phys}}, \delta x_s &\equiv&
  %  \frac{m_{s}^\mathrm{val} - m_{s}^\mathrm{phys}}{m_{s}^\mathrm{phys}}.\nonumber
%\end{eqnarray}

\vspace{1em}
Discretisation effects are handled by $c_{a^2}$, though negligibly small. The fit from all~10 of our configuration sets is excellent, with a $\chi^2$ per degree of freedom of 0.22 ($p$-value of 0.99). 

Our final result for the connected contribution for $s$~quarks to $g-2$ is (figure~\ref{fig:s-fit}):
\begin{equation}
    a_\mu^s = 53.41(59) \times 10^{-10}.
\end{equation}
The precision obtained is at the 1.1$\%$ level where the lattice spacing uncertainty alone contributes $\sim$ 1$\%$. This can certainly be improved to achieve a better precision. Finite volume effects seemed to be negligibly small. But, the valence HISQ strange quark mass tuning effect was significant. Disconnected diagrams are not included in the calculation, but we expect these to have only a very small effect.

\subsection{Comparison of our results for $a_\mu^s$ and $a_\mu^c$ with other results}

\begin{figure}
   \centering
   \includegraphics[width=6.5cm,angle=0]{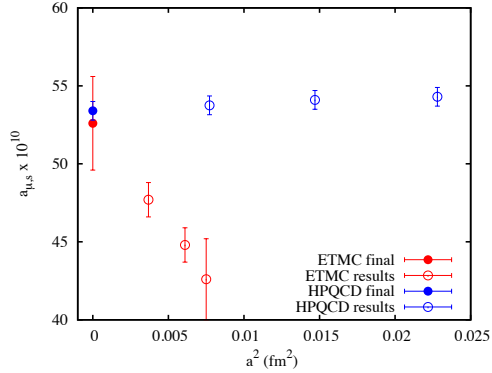}
\caption{Comparison of our results for $a_\mu^s$ with the ETMC preliminary results~\cite{privateETMC}}
\label{fig:comp}
\end{figure}
%{\footnotesize{\begin{itemize}
Our result for $a_\mu^s$ in the continuum limit agrees with the lattice results later obtained by the ETM Collaboration~\cite{privateETMC}. As shown in figure~\ref{fig:comp}, the HISQ formalism gives much smaller discretisation errors making calculations on relatively coarse lattices viable.

The charm quark contribution (connected) to $a_\mu^{HVP,LO}$ has also been calculated from the previously obtained moments~\cite{Donald:2012ga,jpsi} and found to be 14.42(39)$\times$ 10$^{-10}$. This piece could have been extracted with a similar precision of $\sim$ 1$\%$ by improving the calculation of $Z_{V,\overline{c}c}$ following the same procedure for calculating $Z_{V,\overline{s}s}$~\cite{Chakraborty:2014zma}. However, given the small value of $a_\mu^c$, the uncertainty is not significant. Table~\ref{tab:comp} gives a comparison of our results for $a_\mu^s$ and $a_\mu^c$ with ETMC lattice calculations and the existing most accurate other calculations.
%{\color{bblue}{\begin{equation}
 %   a_\mu^c = 14.42(39)\times 10^{-10}.\nonumber
%\end{equation}}}
%\item Need to precisely calculate $Z_{V,\overline{c}c}$ in the same way as before (B.Chakraborty et al., PoS LATTICE2013, 309(2013)).
%\hspace{-2em}
%\vspace{2.0em}
\begin{table*}
%\tiny
\centering
\caption{Comparison of our results for $a_\mu^s$ and $a_\mu^c$ with ETMC lattice calculations and the results using the dispersion relation and the experimental results on $e^+e^- \rightarrow$ hadrons or $\tau$ decay.}
\label{tab:comp}
\begin{tabular}{cccc}
\hline
\hline
$a_\mu^{s/c}$  & Results from dispersion   &  Our results  & ETMC (preliminary) \\
               & + experiment              &  ~\cite{PhysRevD.89.114501}             & results~\cite{privateETMC}\\
\hline
$a_\mu^s$  & 55.3(8)x10$^{-10}$~\cite{Hagiwara:2011af,PhysRevD.89.114501}         & 53.41(59)x10$^{-10}$  &  53(3)x10$^{-10}$ \\
\hline
$a_\mu^c$  & 14.4(1)x10$^{-10}$~\cite{Bodenstein:2011qy}       & 14.42(39)x10$^{-10}$               & 14.1(6)x10$^{-10}$ \\
\hline
\hline
\end{tabular}%}}%}
\end{table*}%}}

\subsection{Preliminary results of the connected contribution to $a_{\mu}^{light}$}

The light (up and down) quark contribution in $a_\mu^{HVP,LO}$ is the most significant part, being 12 times larger than that for the strange quark, in part beacuse of a factor of 5 from the electric charges. Though the extension of our method to calculate $a_{\mu}^{light}$ is straightforward, poor signal-to-noise ratio in this case makes the calculation challenging by significantly increasing the statistical errors in the moments. We have overcome this issue by calculating the time moments from the reconstructed correlators using the best fit parameters (instead of using the original correlators). This constrains the errors in the correlators at larger times therefore gives a much better precision.
%\begin{figure}
 %  \centering                                                                                                                              
  % \includegraphics[width=5.0cm,angle=0]{amu.pdf}
  % \caption{The $a_{\mu}^q$ for different valence quarks (masses) are shown on very coarse and coarse ensembles.}
  % \label{fig:amu_l}
%\end{figure} 

We have achieved a 5-6$\%$ uncertainty (preliminary) in $a_{\mu}^{light}$ using 12 time sources on each of the 1000 configurations on very coarse lattices and only 4 time sources on each of the 400 configurations on coarse (physical point) lattices. We are currently improving these results with higher statistics. 
%To achieve 1$\%$ precision, need 4 x time sources and up to 10 x configurations.

We estimate the total $a_{\mu}^{HVP,LO} = a_{\mu}^{light}+a_{\mu}^s+a_{\mu}^c \sim$ 662(35)x10$^{-10}$ (by averaging $a_{\mu}^{light}$ on physical point coarse and very coarse ensembles).

\section{Conclusion and Ongoing Work}

To summarize, we have developed a simple lattice QCD method and, using that method, achieved 1$\%$ uncertainty for the connected strange quark contribution to $a_{\mu}^{HVP,LO}$ along with an accurate result for $a_{\mu}^c$. The preliminary calculation of connected light quark contributions to $a_{\mu}^{HVP,LO}$ gives a 5-6$\%$ uncertainty on physical point lattices, but without yet using the full statistics currently available. 

Our ongoing work using the full set of configurations, more time sources on the existing configurations and smearing aims to achieve a few percent uncertainty in $a_{\mu}^{light}$. More configurations can also be made for very coarse and coarse lattices relatively cheaply. The estimated (preliminary) total $a_{\mu}^{HVP,LO}$ is $\sim$ 662(35)x10$^{-10}$, but the disconnected contribution still needs to be included. We are optimistic about obtaining an uncertainty in $a_{\mu}^{HVP,LO}$ of less than 1$\%$ level during 2015.

\subsection*{\bf{Acknowledgements}}
 We are grateful to the MILC collaboration for the use of their gauge configurations. We thank R.~R.~Horgan and G.~M.~von~Hippel for conversations. Our calculations were done on the Darwin Supercomputer as part of STFC's DiRAC facility jointly funded by STFC, BIS and the Universities of Cambridge and Glasgow. This work was funded by STFC, the Royal Society, the Wolfson Foundation and the National Science Foundation.

\bibliographystyle{h-physrev.bst}
\bibliography{g2s}{}

\begin{thebibliography}{30}
\expandafter\ifx\csname natexlab\endcsname\relax\def\natexlab#1{#1}\fi
\expandafter\ifx\csname bibnamefont\endcsname\relax
  \def\bibnamefont#1{#1}\fi
\expandafter\ifx\csname bibfnamefont\endcsname\relax
  \def\bibfnamefont#1{#1}\fi
\expandafter\ifx\csname citenamefont\endcsname\relax
  \def\citenamefont#1{#1}\fi
\expandafter\ifx\csname url\endcsname\relax
  \def\url#1{\texttt{#1}}\fi
\expandafter\ifx\csname urlprefix\endcsname\relax\def\urlprefix{URL }\fi
\providecommand{\bibinfo}[2]{#2}
\providecommand{\eprint}[2][]{\url{#2}}

\bibitem[{\citenamefont{Bennett et~al.}(2006)}]{Bennett:2006fi}
\bibinfo{author}{\bibfnamefont{G.}~\bibnamefont{Bennett}} \bibnamefont{et~al.}
  (\bibinfo{collaboration}{Muon G-2 Collaboration}),
  \bibinfo{journal}{Phys.Rev.} \textbf{\bibinfo{volume}{D73}},
  \bibinfo{pages}{072003} (\bibinfo{year}{2006}), \eprint{hep-ex/0602035}.

\bibitem[{\citenamefont{Aoyama et~al.}(2012)\citenamefont{Aoyama, Hayakawa,
  Kinoshita, and Nio}}]{Aoyama:2012wk}
\bibinfo{author}{\bibfnamefont{T.}~\bibnamefont{Aoyama}},
  \bibinfo{author}{\bibfnamefont{M.}~\bibnamefont{Hayakawa}},
  \bibinfo{author}{\bibfnamefont{T.}~\bibnamefont{Kinoshita}},
  \bibnamefont{and} \bibinfo{author}{\bibfnamefont{M.}~\bibnamefont{Nio}},
  \bibinfo{journal}{Phys.Rev.Lett.} \textbf{\bibinfo{volume}{109}},
  \bibinfo{pages}{111808} (\bibinfo{year}{2012}), \eprint{1205.5370}.

\bibitem[{\citenamefont{Hagiwara et~al.}(2011)\citenamefont{Hagiwara, Liao,
  Martin, Nomura, and Teubner}}]{Hagiwara:2011af}
\bibinfo{author}{\bibfnamefont{K.}~\bibnamefont{Hagiwara}},
  \bibinfo{author}{\bibfnamefont{R.}~\bibnamefont{Liao}},
  \bibinfo{author}{\bibfnamefont{A.~D.} \bibnamefont{Martin}},
  \bibinfo{author}{\bibfnamefont{D.}~\bibnamefont{Nomura}}, \bibnamefont{and}
  \bibinfo{author}{\bibfnamefont{T.}~\bibnamefont{Teubner}},
  \bibinfo{journal}{J.Phys.} \textbf{\bibinfo{volume}{G38}},
  \bibinfo{pages}{085003} (\bibinfo{year}{2011}), \eprint{1105.3149}.

\bibitem[{\citenamefont{Davier et~al.}(2011)\citenamefont{Davier, Hoecker,
  Malaescu, and Zhang}}]{Davier}
\bibinfo{author}{\bibfnamefont{M.}~\bibnamefont{Davier}},
  \bibinfo{author}{\bibfnamefont{A.}~\bibnamefont{Hoecker}},
  \bibinfo{author}{\bibfnamefont{B.}~\bibnamefont{Malaescu}}, \bibnamefont{and}
  \bibinfo{author}{\bibfnamefont{Z.}~\bibnamefont{Zhang}},
  \bibinfo{journal}{Eur.Phys.J.} \textbf{\bibinfo{volume}{C71}},
  \bibinfo{pages}{1515} (\bibinfo{year}{2011}), \eprint{1010.4180}.

\bibitem[{\citenamefont{Blum}(2003)}]{Blum:2002ii}
\bibinfo{author}{\bibfnamefont{T.}~\bibnamefont{Blum}},
  \bibinfo{journal}{Phys.Rev.Lett.} \textbf{\bibinfo{volume}{91}},
  \bibinfo{pages}{052001} (\bibinfo{year}{2003}), \eprint{hep-lat/0212018}.

\bibitem[{\citenamefont{Blum et~al.}(2012)\citenamefont{Blum, Hayakawa, and
  Izubuchi}}]{Blum:2013qu}
\bibinfo{author}{\bibfnamefont{T.}~\bibnamefont{Blum}},
  \bibinfo{author}{\bibfnamefont{M.}~\bibnamefont{Hayakawa}}, \bibnamefont{and}
  \bibinfo{author}{\bibfnamefont{T.}~\bibnamefont{Izubuchi}},
  \bibinfo{journal}{PoS} \textbf{\bibinfo{volume}{LATTICE2012}},
  \bibinfo{pages}{022} (\bibinfo{year}{2012}), \eprint{1301.2607}.

\bibitem[{\citenamefont{Aubin and Blum}(2007)}]{aubin-blum}
\bibinfo{author}{\bibfnamefont{C.}~\bibnamefont{Aubin}} \bibnamefont{and}
  \bibinfo{author}{\bibfnamefont{T.}~\bibnamefont{Blum}},
  \bibinfo{journal}{Phys.Rev.} \textbf{\bibinfo{volume}{D75}},
  \bibinfo{pages}{114502} (\bibinfo{year}{2007}), \eprint{hep-lat/0608011}.

\bibitem[{\citenamefont{Boyle et~al.}(2012)\citenamefont{Boyle, Del~Debbio,
  Kerrane, and Zanotti}}]{boyle}
\bibinfo{author}{\bibfnamefont{P.}~\bibnamefont{Boyle}},
  \bibinfo{author}{\bibfnamefont{L.}~\bibnamefont{Del~Debbio}},
  \bibinfo{author}{\bibfnamefont{E.}~\bibnamefont{Kerrane}}, \bibnamefont{and}
  \bibinfo{author}{\bibfnamefont{J.}~\bibnamefont{Zanotti}},
  \bibinfo{journal}{Phys.Rev.} \textbf{\bibinfo{volume}{D85}},
  \bibinfo{pages}{074504} (\bibinfo{year}{2012}), \eprint{1107.1497}.

\bibitem[{\citenamefont{Feng et~al.}(2011)\citenamefont{Feng, Jansen,
  Petschlies, and Renner}}]{renner}
\bibinfo{author}{\bibfnamefont{X.}~\bibnamefont{Feng}},
  \bibinfo{author}{\bibfnamefont{K.}~\bibnamefont{Jansen}},
  \bibinfo{author}{\bibfnamefont{M.}~\bibnamefont{Petschlies}},
  \bibnamefont{and} \bibinfo{author}{\bibfnamefont{D.~B.}
  \bibnamefont{Renner}}, \bibinfo{journal}{Phys.Rev.Lett.}
  \textbf{\bibinfo{volume}{107}}, \bibinfo{pages}{081802}
  (\bibinfo{year}{2011}), \eprint{1103.4818}.

\bibitem[{\citenamefont{Della~Morte et~al.}(2012)\citenamefont{Della~Morte,
  Jager, Juttner, and Wittig}}]{wittig}
\bibinfo{author}{\bibfnamefont{M.}~\bibnamefont{Della~Morte}},
  \bibinfo{author}{\bibfnamefont{B.}~\bibnamefont{Jager}},
  \bibinfo{author}{\bibfnamefont{A.}~\bibnamefont{Juttner}}, \bibnamefont{and}
  \bibinfo{author}{\bibfnamefont{H.}~\bibnamefont{Wittig}},
  \bibinfo{journal}{JHEP} \textbf{\bibinfo{volume}{1203}}, \bibinfo{pages}{055}
  (\bibinfo{year}{2012}), \eprint{1112.2894}.

\bibitem[{\citenamefont{Aubin et~al.}(2013)\citenamefont{Aubin, Blum,
  Golterman, and Peris}}]{aubin-twisted}
\bibinfo{author}{\bibfnamefont{C.}~\bibnamefont{Aubin}},
  \bibinfo{author}{\bibfnamefont{T.}~\bibnamefont{Blum}},
  \bibinfo{author}{\bibfnamefont{M.}~\bibnamefont{Golterman}},
  \bibnamefont{and} \bibinfo{author}{\bibfnamefont{S.}~\bibnamefont{Peris}},
  \bibinfo{journal}{Phys.Rev.} \textbf{\bibinfo{volume}{D88}},
  \bibinfo{pages}{074505} (\bibinfo{year}{2013}), \eprint{1307.4701}.

\bibitem[{\citenamefont{Aubin et~al.}(2012)\citenamefont{Aubin, Blum,
  Golterman, and Peris}}]{golterman-pade}
\bibinfo{author}{\bibfnamefont{C.}~\bibnamefont{Aubin}},
  \bibinfo{author}{\bibfnamefont{T.}~\bibnamefont{Blum}},
  \bibinfo{author}{\bibfnamefont{M.}~\bibnamefont{Golterman}},
  \bibnamefont{and} \bibinfo{author}{\bibfnamefont{S.}~\bibnamefont{Peris}},
  \bibinfo{journal}{PoS} \textbf{\bibinfo{volume}{LATTICE2012}},
  \bibinfo{pages}{176} (\bibinfo{year}{2012}), \eprint{1210.7611}.

\bibitem[{Pad()}]{Pade}
\bibinfo{note}{{The $[m,n]$ Pad\'e approximant of a function $f(x)$ is a ratio
  of polynomials in~$x$, of order $m$ in the numerator and $n$ in the
  denominator, whose Taylor expansion is the same as that of $f(x)$ through
  order~$x^{n+m}$.}}

\bibitem[{\citenamefont{Baker}(1969)}]{Baker:1969}
\bibinfo{author}{\bibfnamefont{G.}~\bibnamefont{Baker}},
  \bibinfo{journal}{J.Math.Phys.} \textbf{\bibinfo{volume}{10}},
  \bibinfo{pages}{814} (\bibinfo{year}{1969}),
  \eprint{http://dx.doi.org/10.1063/1.1664911}.

\bibitem[{\citenamefont{Baker and Graves-Morris}(1996)}]{Baker:1996}
\bibinfo{author}{\bibfnamefont{G.}~\bibnamefont{Baker}} \bibnamefont{and}
  \bibinfo{author}{\bibfnamefont{P.}~\bibnamefont{Graves-Morris}},
  \emph{\bibinfo{title}{Pad\'e Approximants}} (\bibinfo{publisher}{Cambridge},
  \bibinfo{year}{1996}), \bibinfo{edition}{2nd} ed., \bibinfo{note}{especially
  chapter~5}.

\bibitem[{\citenamefont{Donald et~al.}(2012)\citenamefont{Donald, Davies,
  Dowdall, Follana, Hornbostel et~al.}}]{Donald:2012ga}
\bibinfo{author}{\bibfnamefont{G.}~\bibnamefont{Donald}},
  \bibinfo{author}{\bibfnamefont{C.}~\bibnamefont{Davies}},
  \bibinfo{author}{\bibfnamefont{R.}~\bibnamefont{Dowdall}},
  \bibinfo{author}{\bibfnamefont{E.}~\bibnamefont{Follana}},
  \bibinfo{author}{\bibfnamefont{K.}~\bibnamefont{Hornbostel}},
  \bibnamefont{et~al.}, \bibinfo{journal}{Phys.Rev.}
  \textbf{\bibinfo{volume}{D86}}, \bibinfo{pages}{094501}
  (\bibinfo{year}{2012}), \eprint{1208.2855}.

\bibitem[{\citenamefont{Davies et~al.}(2012)\citenamefont{Davies, Donald,
  Dowdall, Koponen, Follana et~al.}}]{jpsi}
\bibinfo{author}{\bibfnamefont{C.}~\bibnamefont{Davies}},
  \bibinfo{author}{\bibfnamefont{G.}~\bibnamefont{Donald}},
  \bibinfo{author}{\bibfnamefont{R.}~\bibnamefont{Dowdall}},
  \bibinfo{author}{\bibfnamefont{J.}~\bibnamefont{Koponen}},
  \bibinfo{author}{\bibfnamefont{E.}~\bibnamefont{Follana}},
  \bibnamefont{et~al.}, \bibinfo{journal}{PoS}
  \textbf{\bibinfo{volume}{ConfinementX}}, \bibinfo{pages}{288}
  (\bibinfo{year}{2012}), \eprint{1301.7203}.

\bibitem[{\citenamefont{Bazavov et~al.}(2010)}]{Bazavov:2010ru}
\bibinfo{author}{\bibfnamefont{A.}~\bibnamefont{Bazavov}} \bibnamefont{et~al.}
  (\bibinfo{collaboration}{MILC collaboration}), \bibinfo{journal}{Phys.Rev.}
  \textbf{\bibinfo{volume}{D82}}, \bibinfo{pages}{074501}
  (\bibinfo{year}{2010}), \eprint{1004.0342}.

\bibitem[{\citenamefont{Bazavov et~al.}(2013)}]{Bazavov:2012uw}
\bibinfo{author}{\bibfnamefont{A.}~\bibnamefont{Bazavov}} \bibnamefont{et~al.}
  (\bibinfo{collaboration}{MILC Collaboration}), \bibinfo{journal}{Phys.Rev.}
  \textbf{\bibinfo{volume}{D87}}, \bibinfo{pages}{054505}
  (\bibinfo{year}{2013}), \eprint{1212.4768}.

\bibitem[{\citenamefont{Dowdall et~al.}(2013)\citenamefont{Dowdall, Davies,
  Lepage, and McNeile}}]{fkpi}
\bibinfo{author}{\bibfnamefont{R.}~\bibnamefont{Dowdall}},
  \bibinfo{author}{\bibfnamefont{C.}~\bibnamefont{Davies}},
  \bibinfo{author}{\bibfnamefont{G.}~\bibnamefont{Lepage}}, \bibnamefont{and}
  \bibinfo{author}{\bibfnamefont{C.}~\bibnamefont{McNeile}},
  \bibinfo{journal}{Phys.Rev.} \textbf{\bibinfo{volume}{D88}},
  \bibinfo{pages}{074504} (\bibinfo{year}{2013}), \eprint{1303.1670}.

\bibitem[{\citenamefont{Borsanyi et~al.}(2012)\citenamefont{Borsanyi, Durr,
  Fodor, Hoelbling, Katz et~al.}}]{Borsanyi:2012zs}
\bibinfo{author}{\bibfnamefont{S.}~\bibnamefont{Borsanyi}},
  \bibinfo{author}{\bibfnamefont{S.}~\bibnamefont{Durr}},
  \bibinfo{author}{\bibfnamefont{Z.}~\bibnamefont{Fodor}},
  \bibinfo{author}{\bibfnamefont{C.}~\bibnamefont{Hoelbling}},
  \bibinfo{author}{\bibfnamefont{S.~D.} \bibnamefont{Katz}},
  \bibnamefont{et~al.}, \bibinfo{journal}{JHEP}
  \textbf{\bibinfo{volume}{1209}}, \bibinfo{pages}{010} (\bibinfo{year}{2012}),
  \eprint{1203.4469}.

\bibitem[{\citenamefont{Chakraborty et~al.}(2013)\citenamefont{Chakraborty,
  Davies, Donald, Dowdall, Koponen et~al.}}]{Chakraborty:2014zma}
\bibinfo{author}{\bibfnamefont{B.}~\bibnamefont{Chakraborty}},
  \bibinfo{author}{\bibfnamefont{C.}~\bibnamefont{Davies}},
  \bibinfo{author}{\bibfnamefont{G.}~\bibnamefont{Donald}},
  \bibinfo{author}{\bibfnamefont{R.}~\bibnamefont{Dowdall}},
  \bibinfo{author}{\bibfnamefont{J.}~\bibnamefont{Koponen}},
  \bibnamefont{et~al.}, \bibinfo{journal}{PoS}
  \textbf{\bibinfo{volume}{LATTICE2013}}, \bibinfo{pages}{309}
  (\bibinfo{year}{2013}), \eprint{1401.0669}.

\bibitem[{\citenamefont{Follana et~al.}(2007)\citenamefont{Follana, Mason,
  Davies, Hornbostel, Lepage et~al.}}]{HISQ_PRD}
\bibinfo{author}{\bibfnamefont{E.}~\bibnamefont{Follana}},
  \bibinfo{author}{\bibfnamefont{Q.}~\bibnamefont{Mason}},
  \bibinfo{author}{\bibfnamefont{C.}~\bibnamefont{Davies}},
  \bibinfo{author}{\bibfnamefont{K.}~\bibnamefont{Hornbostel}},
  \bibinfo{author}{\bibfnamefont{G.~P.} \bibnamefont{Lepage}},
  \bibnamefont{et~al.} (\bibinfo{collaboration}{HPQCD and UKQCD
  Collaborations}), \bibinfo{journal}{Phys.Rev.}
  \textbf{\bibinfo{volume}{D75}}, \bibinfo{pages}{054502}
  (\bibinfo{year}{2007}), \eprint{hep-lat/0610092}.

\bibitem[{\citenamefont{Follana et~al.}(2008)\citenamefont{Follana, Davies,
  Lepage, and Shigemitsu}}]{HISQ_PRL}
\bibinfo{author}{\bibfnamefont{E.}~\bibnamefont{Follana}},
  \bibinfo{author}{\bibfnamefont{C.}~\bibnamefont{Davies}},
  \bibinfo{author}{\bibfnamefont{G.}~\bibnamefont{Lepage}}, \bibnamefont{and}
  \bibinfo{author}{\bibfnamefont{J.}~\bibnamefont{Shigemitsu}}
  (\bibinfo{collaboration}{HPQCD and UKQCD Collaborations}),
  \bibinfo{journal}{Phys.Rev.Lett.} \textbf{\bibinfo{volume}{100}},
  \bibinfo{pages}{062002} (\bibinfo{year}{2008}), \eprint{0706.1726}.

\bibitem[{\citenamefont{Davies et~al.}(2010{\natexlab{a}})\citenamefont{Davies,
  McNeile, Follana, Lepage, Na et~al.}}]{Dsdecayconst}
\bibinfo{author}{\bibfnamefont{C.}~\bibnamefont{Davies}},
  \bibinfo{author}{\bibfnamefont{C.}~\bibnamefont{McNeile}},
  \bibinfo{author}{\bibfnamefont{E.}~\bibnamefont{Follana}},
  \bibinfo{author}{\bibfnamefont{G.}~\bibnamefont{Lepage}},
  \bibinfo{author}{\bibfnamefont{H.}~\bibnamefont{Na}}, \bibnamefont{et~al.}
  (\bibinfo{collaboration}{HPQCD Collaboration}), \bibinfo{journal}{Phys.Rev.}
  \textbf{\bibinfo{volume}{D82}}, \bibinfo{pages}{114504}
  (\bibinfo{year}{2010}{\natexlab{a}}), \eprint{1008.4018}.

\bibitem[{\citenamefont{Durr et~al.}(2011)\citenamefont{Durr, Fodor, Hoelbling,
  Katz, Krieg et~al.}}]{durrmsml}
\bibinfo{author}{\bibfnamefont{S.}~\bibnamefont{Durr}},
  \bibinfo{author}{\bibfnamefont{Z.}~\bibnamefont{Fodor}},
  \bibinfo{author}{\bibfnamefont{C.}~\bibnamefont{Hoelbling}},
  \bibinfo{author}{\bibfnamefont{S.}~\bibnamefont{Katz}},
  \bibinfo{author}{\bibfnamefont{S.}~\bibnamefont{Krieg}},
  \bibnamefont{et~al.}, \bibinfo{journal}{Phys.Lett.}
  \textbf{\bibinfo{volume}{B701}}, \bibinfo{pages}{265} (\bibinfo{year}{2011}),
  \eprint{1011.2403}.

\bibitem[{\citenamefont{Davies et~al.}(2010{\natexlab{b}})\citenamefont{Davies,
  Follana, Kendall, Lepage, and McNeile}}]{Davies:2009tsa}
\bibinfo{author}{\bibfnamefont{C.}~\bibnamefont{Davies}},
  \bibinfo{author}{\bibfnamefont{E.}~\bibnamefont{Follana}},
  \bibinfo{author}{\bibfnamefont{I.}~\bibnamefont{Kendall}},
  \bibinfo{author}{\bibfnamefont{G.}~\bibnamefont{Lepage}}, \bibnamefont{and}
  \bibinfo{author}{\bibfnamefont{C.}~\bibnamefont{McNeile}}
  (\bibinfo{collaboration}{HPQCD Collaboration}), \bibinfo{journal}{Phys.Rev.}
  \textbf{\bibinfo{volume}{D81}}, \bibinfo{pages}{034506}
  (\bibinfo{year}{2010}{\natexlab{b}}), \eprint{0910.1229}.

\bibitem[{\citenamefont{Allison et~al.}(2008)}]{Allison:2008xk}
\bibinfo{author}{\bibfnamefont{I.}~\bibnamefont{Allison}} \bibnamefont{et~al.}
  (\bibinfo{collaboration}{HPQCD Collaboration}), \bibinfo{journal}{Phys.Rev.}
  \textbf{\bibinfo{volume}{D78}}, \bibinfo{pages}{054513}
  (\bibinfo{year}{2008}), \eprint{0805.2999}.

\bibitem[{\citenamefont{McNeile et~al.}(2010)\citenamefont{McNeile, Davies,
  Follana, Hornbostel, and Lepage}}]{McNeile:2010ji}
\bibinfo{author}{\bibfnamefont{C.}~\bibnamefont{McNeile}},
  \bibinfo{author}{\bibfnamefont{C.}~\bibnamefont{Davies}},
  \bibinfo{author}{\bibfnamefont{E.}~\bibnamefont{Follana}},
  \bibinfo{author}{\bibfnamefont{K.}~\bibnamefont{Hornbostel}},
  \bibnamefont{and} \bibinfo{author}{\bibfnamefont{G.}~\bibnamefont{Lepage}},
  \bibinfo{journal}{Phys.Rev.} \textbf{\bibinfo{volume}{D82}},
  \bibinfo{pages}{034512} (\bibinfo{year}{2010}), \eprint{1004.4285}.

\bibitem[{\citenamefont{Gonnet et~al.}(2013)}]{Gonnet:2013}
\bibinfo{author}{\bibfnamefont{P.}~\bibnamefont{Gonnet}} \bibnamefont{et~al.},
  \bibinfo{journal}{SIAM Review} \textbf{\bibinfo{volume}{55}},
  \bibinfo{pages}{101} (\bibinfo{year}{2013}).

\end{thebibliography}


\begin{thebibliography}{10}

\bibitem{Bennett:2006fi}
Muon G-2 Collaboration, G.~Bennett {\em et~al.},
\newblock Phys.Rev. {\bf D73}, 072003 (2006), hep-ex/0602035.

\bibitem{Aoyama:2012wk}
T.~Aoyama, M.~Hayakawa, T.~Kinoshita, and M.~Nio,
\newblock Phys.Rev.Lett. {\bf 109}, 111808 (2012), 1205.5370.

\bibitem{Hagiwara:2011af}
K.~Hagiwara, R.~Liao, A.~D. Martin, D.~Nomura, and T.~Teubner,
\newblock J.Phys. {\bf G38}, 085003 (2011), 1105.3149.

\bibitem{rafael1972}
{B.E. Lautrup, A. Peterman and E. de Rafael, Phys. Rep. 3, N$^{\circ}$4 (1972)
  193-260}.

\bibitem{Davier}
M.~Davier, A.~Hoecker, B.~Malaescu, and Z.~Zhang,
\newblock Eur.Phys.J. {\bf C71}, 1515 (2011), 1010.4180.

\bibitem{PhysRevD.89.114501}
HPQCD Collaboration, B.~Chakraborty {\em et~al.},
\newblock Phys. Rev. D {\bf 89}, 114501 (2014).

\bibitem{rafael1994}
{E. de Rafael, Phys. Lett; B 322 (1994) 239}.

\bibitem{Blum:2002ii}
T.~Blum,
\newblock Phys.Rev.Lett. {\bf 91}, 052001 (2003), hep-lat/0212018.

\bibitem{Allison:2008xk}
HPQCD Collaboration, I.~Allison {\em et~al.},
\newblock Phys.Rev. {\bf D78}, 054513 (2008), 0805.2999.

\bibitem{McNeile:2010ji}
HPQCD Collaboration, C.~McNeile, C.~Davies, E.~Follana, K.~Hornbostel, and
  G.~Lepage,
\newblock Phys.Rev. {\bf D82}, 034512 (2010), 1004.4285.

\bibitem{Pade}
{The $[m,n]$ Pad\'e approximant of a function $f(x)$ is a ratio of polynomials
  in~$x$, of order $m$ in the numerator and $n$ in the denominator, whose
  Taylor expansion is the same as that of $f(x)$ through order~$x^{n+m}$.}

\bibitem{HISQ_PRD}
HPQCD and UKQCD Collaborations, E.~Follana {\em et~al.},
\newblock Phys.Rev. {\bf D75}, 054502 (2007), hep-lat/0610092.

\bibitem{Bazavov:2010ru}
MILC collaboration, A.~Bazavov {\em et~al.},
\newblock Phys.Rev. {\bf D82}, 074501 (2010), 1004.0342.

\bibitem{Bazavov:2012uw}
MILC Collaboration, A.~Bazavov {\em et~al.},
\newblock Phys.Rev. {\bf D87}, 054505 (2013), 1212.4768.

\bibitem{fkpi}
HPQCD Collaboration, R.~Dowdall, C.~Davies, G.~Lepage, and C.~McNeile,
\newblock Phys.Rev. {\bf D88}, 074504 (2013), 1303.1670.

\bibitem{Borsanyi:2012zs}
S.~Borsanyi {\em et~al.},
\newblock JHEP {\bf 1209}, 010 (2012), 1203.4469.

\bibitem{Chakraborty:2014zma}
HPQCD and UKQCD Collaborations, B.~Chakraborty {\em et~al.},
\newblock PoS {\bf LATTICE2013}, 309 (2013), 1401.0669.

\bibitem{privateETMC}
{Private communication with ETM Collaboration about their preliminary results}.

\bibitem{Donald:2012ga}
HPQCD Collaboration, G.~Donald {\em et~al.},
\newblock Phys.Rev. {\bf D86}, 094501 (2012), 1208.2855.

\bibitem{jpsi}
HPQCD Collaboration, C.~Davies {\em et~al.},
\newblock PoS {\bf ConfinementX}, 288 (2012), 1301.7203.

\bibitem{Bodenstein:2011qy}
S.~Bodenstein, C.~Dominguez, and K.~Schilcher,
\newblock Phys.Rev. {\bf D85}, 014029 (2012), 1106.0427.

\end{thebibliography}

\end{document}